\documentstyle[epsf]{aa}
\begin{document}
\def\gtsima{$\; \buildrel > \over \sim \;$}
\def\simgt{\lower.5ex\hbox{\gtsima}}
\thesaurus{12(12.03.1,12.03.3, 12.07.1)}
\title{Moving gravitational lenses: imprints on the CMB}
\author{N. Aghanim\inst{1} \and S. Prunet\inst{1} \and O. Forni\inst{1} 
\and F. R. Bouchet\inst{2}}
\offprints{N. Aghanim }

\institute{IAS-CNRS, Universit\'e Paris XI, B\^atiment 121, F-91405 Orsay Cedex
\and IAP-CNRS, 98 bis, Boulevard Arago, F-75014 Paris}

\date{Received date / accepted date}
\maketitle
\begin{abstract}
With the new generation of instruments for Cosmic Microwave Background 
(CMB) observations aiming at 
an accuracy level of a few percent in the measurement 
of 
the angular power spectrum of the anisotropies, the study of the contributions 
due to secondary
effects has gained impetus. Furthermore,
a reinvestigation of the
main secondary effects is crucial in order to predict and quantify their 
effects on the CMB and the errors that they induce in the measurements. 
\par
In this paper, we investigate the contribution, to the CMB, of secondary 
anisotropies induced by the transverse motions of clusters of
galaxies. This effect is similar
to the Kaiser--Stebbins effect. In order to address this problem,
we model the gravitational potential well of an individual structure using the
Navarro, Frenk \& White profile. We generalise the effect of one structure to
a population of objects predicted using the Press-Schechter formalism. We
simulate maps of these secondary fluctuations, compute the angular power 
spectrum and derive the average contributions for three cosmological models.
We then investigate a simple method to separate this new contribution
from the primary anisotropies and from the main 
secondary effect, the Sunyaev-Zel'dovich kinetic effect 
from the lensing
clusters.
\par
\end{abstract}
\section{Introduction}
During the next decade, several experiments are planned to 
observe 
the 
Cosmic Microwave Background (CMB) and measure its temperature fluctuations  
(Planck surveyor, Map, Boomerang, ...). Their challenge is to measure
the small scales anisotropies of 
the CMB (a few arcminutes up to ten degrees scale) with 
sensitivities better by a factor 10 than the COBE satellite (Smoot et al. 
1992). These high sensitivity and resolution measurements 
will tightly  
constrain the value of the main cosmological parameters (Kamionkowski et al. 
1994). However, the constraints can only be set if 
we are able to effectively
measure the {\it primary} temperature fluctuations. These fluctuations, 
present at recombination, give an insight  into the early universe since they are 
directly related to the initial 
density perturbations which are the progenitors to
the cosmic structures (galaxies and galaxies clusters) in the present universe; 
but which are first and foremost the 
relics of the very early initial conditions of the universe.\\ 
Between recombination and the present time, the CMB photons 
could have undergone various interactions with the matter and structures 
present along their lines of sight. Some of these 
interactions can induce additional temperature fluctuations called, 
{\it secondary} anisotropies because they are generated after the 
recombination. Along a line of sight, one measures temperature fluctuations
which are the superposition of the {\it primary } and {\it secondary} 
anisotropies. As a result, and in the context of the future CMB 
experiments, accurate analysis of the data will be needed in order to account for the 
foreground contributions due to the secondary fluctuations. 
Photon--matter interactions between recombination and the present time 
are due to the
presence of ionised matter or to variations of the gravitational potential 
wells along the lines of sight.
\par
The CMB photons interact with the ionised matter mainly through Compton
interactions. In fact, after recombination the universe could have been 
re-ionised globally or locally. Global early re-ionisation has 
been widely studied (see Dodelson \& Jubas 1995 for a recent review and 
references therein). Its main 
effect is to either smooth or wipe out some of the primary anisotropies; 
but the interactions of the photons with the matter in 
a fully ionised universe can also give rise to secondary 
anisotropies through the Vishniac effect (Vishniac 1987). This second order 
effect has maximum amplitudes for a very early re-ionisation.
The case of a late inhomogeneous re-ionisation and its imprints on the 
CMB fluctuations has been investigated (Aghanim et al. 1996) and found to be 
rather important. In this case, the secondary anisotropies are due to the bulk
motion of ionised clouds with respect to the CMB frame.
When the re-ionisation is localised in hot ionised intra-cluster media the 
photons interact with the free electrons. The inverse Compton scattering 
between photons and electrons leads to the 
so-called Sunyaev-Zel'dovich (hereafter SZ) effect (Sunyaev \& Zel'dovich1972, 
1980). The Compton distortion due to the motion of the electrons 
in the gas is called the thermal SZ effect. The kinetic SZ effect
is a Doppler distortion due to the peculiar bulk motion of the cluster 
with respect
to the Hubble flow. The SZ thermal effect has the unique property of depressing
the CMB brightness in the Rayleigh-Jeans region and increasing its brightness
above a frequency of about 219 GHz. This frequency dependence makes it rather easy 
to observe and separate from the kinetic SZ effect. In fact, the latter has a 
black body spectrum which makes the spectral confusion between kinetic SZ and
primary fluctuations a serious problem. The SZ
effect has been widely studied for individual clusters and for populations of
clusters. For full reviews on the subject we refer the reader to two major
articles: Rephaeli 1995 and Birkinshaw 1997.
These investigations have clearly shown that the SZ effect in clusters of 
galaxies
provides a powerful tool for cosmology through measurements of the Hubble 
constant, the radial peculiar velocity of clusters and consequently the
large scale velocity fields.
\par
Besides the interactions with the ionised matter, some secondary effects arise
when the CMB photons traverse a varying gravitational potential well. In 
fact, if the gravitational potential well crossed by the photons evolves 
between the time they enter the well and the time they leave it, the delay
between entrance and exit is equivalent to a shift in frequency, 
which induces a temperature anisotropy on the CMB. This effect was first 
studied 
by Rees \& Sciama (1968)  for a potential well growing under its
own gravity. Numerous authors have investigated the potential variations due to
collapsing objects and their effect
on the CMB (Kaiser 1982, Nottale 1984, Martinez-Gonz\'alez, Sanz \& Silk 1990, 
Seljak 1996). Similarly, a 
gravitational potential well moving across the line of
sight is equivalent to a varying potential and will thus imprint secondary fluctuations on the CMB. This effect was first 
studied for one cluster of galaxies by Birkinshaw \& Gull (1983) (Sect. 2).
Kaiser \& Stebbins (1984) and Bouchet, Bennett \& Stebbins (1988) investigated 
a similar effect for moving cosmic
strings. Recent work (Tuluie \& Laguna 1995, Tuluie, Laguna \& Anninos 1996)
based on N-body simulations has pointed out this effect 
in a study of the effect of varying potential on rather large angular 
scales ($\simeq 1^{\circ}$). A discussion of some of these results and a 
comparison with ours will follow in the next sections.
\par
In this paper, following the formalism of Birkinshaw \& Gull (1983) 
and Birkinshaw (1989), we
investigate the contribution of secondary anisotropies due to a population of 
collapsed objects moving across the line of sight, these objects
range from small groups to rich clusters in scale ($10^{13}$ to $10^{15}$
$M_{\sun}$). In 
section 2., we first
study in detail the case of a unique collapsed structure.
We use a structure model to compute in particular
the deflection angle and derive the spatial signature of the moving lens 
effect. We then account (Section 3.)
for the contribution, to the primordial cosmological signal, of the whole 
population of collapsed objects
using predicted counts and we simulate maps of
these secondary anisotropies. In section 4., we analyse the simulated maps and
present our results. We give our conclusions in section 5.
\section{Formalism for an individual moving structure}
One of the first studies of the photon--gravitational potential well
interactions is related to  
the Sachs--Wolfe effect (Sachs \& Wolfe 1967). 
At the recombination time ($z\simeq 1100$) the photons and matter decouple 
while they are in potential wells; the photons are redshifted when they
leave the potential wells. This generates the large angular scale
temperature fluctuations.
\\
Other authors have investigated the effect of time varying potentials
on the CMB photons after the recombination, namely the Rees-Sciama effect
(Rees \& Sciama 1968). If the potential well crossed by the photons evolves
between the time they enter and their exit, the extra-time delay they suffer
changes the temperature of the CMB and induces an additional anisotropy. The
variation of the potential well can have an ``intrinsic'' or a ``kinetic'' 
origin. The first case describes the 
evolution with respect to the background density distribution. The second
case is related to the bulk motion of a gravitational potential well across
the line of sight which mimics a time variation of the potential. 
Photons crossing the leading edge of a structure
will be
redshifted because of the increasing depth of the potential well during their
crossing time; while photons crossing the trailing edge of the same structure
are blueshifted. This results in a characteristic spatial signature for 
the induced anisotropy: a hot-cold temperature spot.
\\
The specific effect of a moving cluster across the sky was first studied 
by Birkinshaw \& Gull (1983) (correction to this paper was made in
Birkinshaw 1989) and it was invoked 
as a method to measure the transverse velocity of massive clusters of galaxies.
These authors found that the transverse motion of a 
cluster across the line of sight induces a frequency shift given by:
\begin{equation}
\frac{\Delta\nu}{\nu}=\beta\gamma \sin\alpha \cos\phi\,\delta({\bf b}).
\label{dnu:eq}
\end{equation} 
Here, $\beta$ is the peculiar velocity in units of the speed of light 
($\beta=v/c$), 
$\gamma$ is the Lorenz factor ($\gamma=(1-\beta^2)^{-1/2}$), $\alpha$ and
$\phi$ are respectively the angle between the peculiar velocity ${\bf v}$ and the 
line of sight of the observer and the azimuthal angle in the plane of the sky,
and $\delta ({\bf b})$ is the deflection angle
due to the gravitational lensing by the cluster at a distance equal
to the impact parameter ${\bf b}$.
This frequency shift induces a brightness variation which in turn can be 
expressed as a secondary temperature fluctuation $\delta T/T$. In their paper,
Birkinshaw \& Gull derived an expression for $\delta T/T$ in the 
Rayleigh-Jeans regime, with some specific assumptions on the 
gravitational potential well associated with the cluster. They assumed 
that the 
matter in the galaxy cluster was homogeneously distributed in an isothermal 
sphere of radius $R$, where $R$ is the characteristic scale of the cluster.
\par
In our paper, we basically follow the same formalism as Birkinshaw \& Gull's
using the corrected expression from Birkinshaw 1989.
We compute the gravitational deflection angle at the impact 
parameter $\delta ({\bf b})$, the corresponding frequency shift and 
then derive the associated temperature fluctuation.
\begin{figure}
\epsfxsize=\columnwidth
\hbox{\epsffile{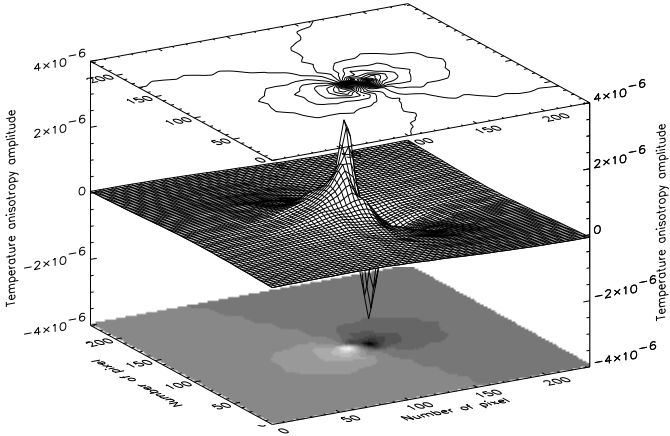}}
\caption{{\small\it Characteristic spatial signature of a temperature 
fluctuation due to a moving lens with mass $M=10^{15}\,M_{\sun}$ and velocity
$v=600$ km/s.}}
\label{fig:farach}
\end{figure}
The main difference between our approach in this section and the previous 
work concerns the
physical hypothesis that we adopt to describe the distribution of 
matter in the structures. In fact, in order to derive the deflection angle, we find 
the homogeneous 
isothermal distribution a too simple and rather unrealistic hypothesis
and choose another more realistic description. For the structures such as those
we are interested in (clusters down to small groups), almost all the mass is
``made'' of dark matter. In order to study the gravitational lensing of a 
structure properly, one has to model the gravitational 
potential well using the best possible knowledge for the dark matter 
distribution.  
The corrections, due to the more accurate profile distribution that we 
introduce, will not alter the
maximum amplitude of an individual moving lens effect since it is associated 
with the 
central part of the lens. However, when dealing with some average signal coming
from 
these secondary anisotropies, the contribution from the outskirts of the
structures appears important and thus a detailed model of the matter profile
is needed. 
\\
In view of the numerous 
recent studies on the formation of dark matter halos, which are the formation
sites for the individual structures such as clusters of galaxies, we 
now have a rather precise idea of  their formation and 
density profiles. Specifically, the results of 
Navarro, Frenk \& White (1996, 1997) are particularly important. In fact, 
these authors have used N-body simulations to investigate the 
structure of dark matter halos in hierarchical cosmogonies; their results 
put stringent constraints on the dark matter profiles. Over about four 
orders of magnitudes in mass (ranging from the masses of dwarf galaxy halos to
those of rich clusters of galaxies), they found that the density profiles can 
be fitted over two decades in radius by a ``universal'' law (hereafter NFW 
profile) which  seems to be the best description of the structure of
dark matter halos (Huss, Jain \& Steinmetz 1997). The NFW profile is given by:
\begin{equation}
\rho(r)=\frac{\rho_{crit}\,\delta_c}{(r/r_s)(1+r/r_s)^2},
\label{nfwprof:eq}
\end{equation}
where $r_s=r_{200}/c$ is the scale radius of the halo, $\delta_c$ its 
characteristic overdensity, $\rho_{crit}$ is the critical density of the
universe and $c$ is a dimensionless parameter called the concentration. 
The radius $r_{200}$ is the radius of the sphere where the mean
density is $200\times\rho_{crit}$. This is what we refer to as a virialised
object of mass $M_{200}=200\rho_{crit}\,(4\pi/3)r_{200}^3$.
\\
In addition to 
the fact that the shape is independent of the halo mass over a wide range, the
NFW profile is also independent of the cosmological model. The cosmological model 
intervenes
essentially in the formation epoch of the dark matter halo and therefore in 
the parameters of the profile, namely $c$, $r_s$ and $\delta_c$. 
\par
Using the density profile, one can compute the deflection angle at 
the impact parameter which gives the shape of the 
pattern and the
amplitude of the induced secondary anisotropy. In our work, we compute the 
deflection angle 
following the formalism of Blandford \& Kochanek (1987), which is given by the 
expression:
%
\begin{equation}
{\bf \delta}=2\frac{D_{ls}}{D_{os}}{\bf \nabla}_{\bf r}\,\int\Phi({\bf r}, l)
\,dl, 
\label{devia:eq}
\end{equation}
here, the integral is performed over the length element $dl$ along the line of 
sight. $D_{ls}$ and $D_{os}$ are respectively the distances between lens and
source and the observer and source. In the redshift range of the
considered structures ($z<1.5$), the distance ratios 
$D_{ls}/D_{os}$ range between 1 and 0.68 for the standard CDM model, between 1
and 0.53 for the open CDM and between 1 and 0.74 for the lambda CDM model. 
These cosmological models will be defined in the next section.
In Eq. \ref{devia:eq} ${\bf r}$ is the position of the structure 
and $\Phi({\bf r}, l)$ is 
the associated gravitational potential. In order to get an analytic expression
of the deflection angle and hence of the anisotropy, we used a density profile 
which gives a good approximation to the NFW density profile (Eq. \ref{nfwprof:eq}), 
in the central part of the structure. This density profile is given by:
\begin{equation}
\rho(r)=\rho_{crit}\,\delta_c\,\left(\frac{r}{r_s}\right)^{-1}
\,\exp\left(\frac{-r}{r_s}\right).
\label{eq:fit}
\end{equation}
The fitted profile leads to a diverging mass at large radii and we therefore 
introduce a cut-off radius 
$R_{max}$ to the integral. This cut-off should correspond to some physical 
size of the structure. With regard to the different values of the 
concentration $c$,
we set $R_{max}= 8r_s$ which is in most cases equivalent to $R_{max}\simeq 
r_{200}$,{\it i.e.,} close to the virial radius. 
The integral 
giving the deflection angle is  
performed on the interval $[-R_{max},R_{max}]$. For $R_{max}= 8r_s$, our fit
gives a mass which is about 20\% lower than the mass derived from NFW profile.
This difference is
larger for larger $R_{max}$, and for
$R_{max}= 10r_s$ we find that the mass is about 33\% lower. 
However, the larger radii
the temperature fluctuations are at the $10^{-8}$ level. On the other 
hand, the Hernquist (1990) profile 
is also in agreement with the results of N-body simulations. Indeed, both NFW 
and Hernquist profiles have a similar dependence
in the central part of the structure but differ at large radii where the
NFW profile is proportional to $r^{-3}$ and the Hernquist profile varies as 
$r^{-4}$. 
However, the amplitude of the anisotropy at large radii is very
small and the results that we obtain does 
are not sensitive to the cut-off.
\par
Given the peculiar velocity of the structure and its density 
profile, we can calculate the deflection angle (Eq. \ref{devia:eq}). 
Then one can determine 
the relative
variation in frequency, $\delta\nu/\nu$, using equation 
\ref{dnu:eq} and thus 
evaluate the secondary distortion induced by a specific structure
moving across the sky. We find that individual massive structures (rich galaxy 
clusters) produce anisotropies ranging between a few $10^{-6}$ to $10^{-5}$; 
but within a wider range of masses the amplitudes are smaller and these values 
are only upper limits for the moving lens effect.
\section{Generalisation to a sample of structures}
Future CMB (space and balloon born) experiments will measure the 
temperature
fluctuations with very high accuracy ($10^{-6}$) at small
angular scales. In our attempts to foresee what the CMB maps would 
look like and what would be the spurious contributions due to the various 
astrophysical foregrounds, we investigate the generalisation of the 
computations made above to a sample of structures. This is done in order to 
address the questions of the cumulative effect and contamination to the CMB.
\par
Some work has already been done by Tuluie \& Laguna 1995 and Tuluie, Laguna \&
Anninos 1996 who pointed out the moving lens effect in their study of 
the varying potential effects on the CMB. In their study, they used N-body 
simulations to evolve
the matter inhomogeneities, from the decoupling time until the present, in which 
they propagated CMB photons. They have estimated the anisotropies generated
by three sources of time--variations of the potential: intrinsic 
changes in the gravitational potential, decaying potential effect from the 
evolution of gravitational potential in $\Omega_0\neq 1$ models, and peculiar 
bulk motions of the
structures across the sky. They evaluated the contribution of
the latter effect for rather large angular scales ($\simeq 1^{\circ}$) due to
the lack of numerical resolution (about $2h^{-1}$ Mpc) and gave estimates of 
the power spectrum of these effects. 
\par
With another approach, we make a similar
analysis in the case of the moving lens effect extended to angular 
scales down to a few tens of 
arcseconds. We also simulate attempts at the detection and 
subtraction of the moving lens effect.
Our approach is quite different from that of Tuluie, Laguna \& 
Anninos, in that it is 
semi-empirical and 
apply the formalism developed for an individual structure (Sect. 2.) 
to each
object from a sample of structures. The predicted number of objects in the 
sample being derived from the Press--Schechter 
formalism for the structure formation (Press \& Schechter 1974).
\subsection{Predicted population of collapsed objects}
An estimate of the cumulative effect of the moving lenses requires  
a knowledge of the number
of objects of a given mass that will contribute to the total effect at a given 
epoch. We assume that this number is accurately predicted by the abundance
of collapsed dark matter halos as a function of their masses and redshifts, as 
derived 
using the Press--Schechter formalism. This approach was used in a previous 
paper (Aghanim et al. 1997) which predicted the SZ contribution to the CMB 
signal in a standard CDM model. In addition to the ``traditional''
standard Cold Dark Matter (CDM) model ($\Omega_0=1$), in this paper
we also address the question 
of a generalised moving lens effect in other cosmological models. We extend
the Press--Schechter formalism to an open CDM model (OCDM) with no 
cosmological constant ($\Omega_0
=0.3$), and also a flat universe with a non zero cosmological constant 
($\Lambda$CDM model) ($\Omega_0=0.3$ and $\Lambda=0.7$). Here $\Omega_0$ is the
density parameter, $\Lambda$ is the
cosmological constant given in units of $3H_0^2$ and $H_0$ is the Hubble 
constant. We take $H_0=100h$ km/s/Mpc, and assume $h=0.5$ throughout the paper.
\\ 
In any case, the general analytic expression for the number density of 
spherical collapsed halos in the mass range $[M,M+dM]$ can be written as
(Lacey \& Cole 1993):
$$
\frac{dn(M,z)}{dM}=-\,\sqrt{\frac{2}{\pi}}\,\frac{\overline{\rho}(z)}{M^2}
\,\frac{d\ln\,\sigma(M)}{d\ln\,M}\,\frac{\delta_{c0}(z)}{\sigma(M)}\,\times 
$$
\begin{equation}
\hspace*{3cm}
\exp\left[-
\frac{\delta_{c0}^2(z)}{2\sigma^2(M)}\right],
\label{pscount:eq}
\end{equation} 
where $\overline{\rho}(z)$ is the mean background density at redshift $z$ and 
$\delta_{c0}(z)$ is the overdensity of a linearly evolving structure. 
The mass variance $\sigma^2(M)$ of the fluctuation spectrum,
filtered on mass scale $M$, is related to the linear power 
spectrum of the initial density fluctuations $P(k)$ through:
$$
\sigma^2(M)=\frac{1}{2\pi^2}\int^{\infty}_0 k^2P(k)W^2(kR)\,dk,
$$
where $W$ is the Fourier transform of the window function over which the 
variance is smoothed (Peebles 1980) and $R$ is the scale associated with mass 
$M$.
In the assumption of a scale--free initial power spectrum with spectral index 
$n$, the variance on mass scale $M$ can be expressed in terms of $\sigma_8$, 
the {\it rms} density fluctuation in sphere of $8h^{-1}$ Mpc size. The 
relationship
between these two quantities is given by
(Mathiesen \& Evrard 
1997):
$$
\sigma(M)=(1.19\Omega_0)^{\alpha}\sigma_8M^{-\alpha},
$$
with $\alpha=(n+3)/6$. It has been shown that $\sigma_8$ varies 
with the cosmological model and in particular with the density parameter 
$\Omega_0$. A general empirical fitting function ($\sigma_8=A\Omega_0^{-B}$)
was derived from a power spectrum normalisation to the cluster abundance 
with a rather good agreement in the values of the parameters $A$ and $B$
(White, Efstathiou \& Frenk 1993, Eke,
Cole \& Frenk 1996, Viana \& Liddle 1996). In our work, we use the ``best
fitting values'' from Viana \& Liddle (1996) which are $A=0.6$ and 
$B=0.36+0.31\Omega_0-0.28\Omega_0^2$ for an open CDM universe 
($\Omega_0<1$ and $\Lambda=0$)
or $B=0.59-0.16\Omega_0+0.06\Omega_0^2$ for a flat universe with a non zero
cosmological constant ($\Omega_0+\Lambda=1$). 
We use $n=-1$ for the spectral index in the cluster mass regime which is the
theoretically predicted value. Some local constraints on the temperature 
abundance of clusters favour $n=-2$ (Henry \& Arnaud 1991, Oukbir, Bartlett 
\& Blanchard 1997) but we did not investigate this case.
\subsection{Peculiar velocities}
On the scale of clusters of galaxies, typically 8$h^{-1}$ Mpc, one can assume 
that
the density fluctuations are in the linear regime. Therefore the fluctuations 
are closely related to the initial conditions from which the structures arise.
In fact, in the assumption of an isotropic Gaussian distribution of the 
initial density perturbations, the initial power spectrum $P(k)$ gives a 
complete description of the velocity field through the three--dimensional 
{\it rms} velocity ($v_{rms}$) predicted by the 
linear gravitational instability for an irrotational field at a given scale
$R$ (Peebles 1993). This velocity is given by:
\begin{equation}
v_{rms}=a(t)\,H\,f(\Omega,\Lambda)\left[\frac{1}{2\pi^2}\int^{\infty}_0 P(k)
W^2(kR)\,dk\right]^{1/2}
\label{vrms:eq}
\end{equation}
where $a(t)$ is the expansion parameter, the Hubble constant $H$ and the 
density parameter
$\Omega$ vary with time (Caroll, Press \& Turner 1992). The function $f(\Omega,\Lambda)$ is accurately 
approximated by $f(\Omega,\Lambda)=\Omega^{0.6}$ (Peebles 1980) even if 
there is a non zero cosmological constant (Lahav et al. 1991).
Furthermore, under the assumptions of linear regime and Gaussian distribution 
of the density fluctuations, the structures move with respect to the global 
Hubble flow with peculiar velocities following a Gaussian distribution
$f(v)=\frac{1}{v_{rms}\sqrt(2\pi)}\exp(\frac{-v^2}{2v_{rms}^2})$ which is fully
described by $v_{rms}$. This prediction is in agreement with numerical 
simulations (Bahcall et al. 1994, Moscardini et al. 1996).
\par
The present observational status of peculiar cluster velocities 
puts few constraints on the cosmological models. Results from the Hudson (1994)
 sample 
using D$_n$-$\sigma$ and IRTF distance estimators give respectively $v_{rms}=
688\pm82$ and $646\pm120$ km/s, a composite sample gives $v_{rms}=725\pm60$ 
km/s
(Moscardini et al 1996). Giovanelli's (1996) sample gives a smaller value,
$v_{rms}=356\pm37$ km/s.
\\
In our paper we compute the three--dimensional 
{\it rms} peculiar velocity on scale 8$h^{-1}$ Mpc (typical virial radius of
a galaxy cluster) using Eq. \ref{vrms:eq} for the three
cosmological models. This is because large scale velocities are 
mostly sensitive to long wavelength density 
fluctuations. This smoothing allows us to get rid of the nonlinear 
effects on small scales but it also tends to underestimate
the peculiar velocities of the smallest objects that we are interested in.
Nevertheless, with regard to the rather important dispersion in the
observational values (320 $<v_{rms}<$ 780 km/s), we use the predicted 
theoretical values, which range between 400 and 500 km/s, and are hence in 
general agreement with the observational data.

\subsection{Simulations}
For each cosmological model, 
we generate a simulated map of the moving lens effect in order to analyse the 
contribution to the signal in terms of temperature fluctuations. The 
simulations
are essentially based on the studies of Aghanim et al. (1997). In the 
following, we describe briefly the main hypothesis 
that we make in simulating the 
maps of the temperature
fluctuations induced by the moving lens effect associated with small groups and
clusters of galaxies ($10^{13}$ and 
$10^{15}\,M_{\sun}$). The predicted number of massive objects is derived from a
distribution of sources using the Press--Schechter formalism 
normalised (Viana \& Liddle 1996) using the X-ray temperature distribution
function derived from Henry \& Arnaud (1991) data. This normalisation has also
been used by Mathiesen \& Evrard (1997) for the ROSAT Brightest Clusters Sample
compiled by Ebeling et al. (1997). 
The position and direction of motion of each object
are random. Their peculiar velocities are also random within
an assumed Gaussian distribution. Here again, the correlations were neglected
because the effect is maximum very close to the central part of the structure 
(about 100 kpc) whereas the correlation length is between 5 and 20 Mpc 
(Bahcall 1988).
The final maps account for the cumulative effect of the moving lenses with
redshifts lower than $z=1.5$. We refer the reader to Aghanim et al. 
(1997) for a detailed description of the simulation. 
\par
In this paper, some changes and improvements have been made to our previous 
study (Aghanim et al. 1997).
In this paper, the predicted source counts (Eq. \ref{pscount:eq}, Sect. 3.1) 
are in agreement with more recent data. They are also adapted to  
the various cosmological models that we have assumed. The standard deviation 
of the peculiar velocity distribution is computed using equation \ref{vrms:eq} 
and is in
reasonable agreement with the data. The advantage of using this equation is 
that the variations with time and cosmology are directly 
handled in the expression. As we pointed out in section 2, the secondary
effects we study here are associated with the whole mass of the structure, not 
only the gas mass. Therefore, the gas part of structures are modelled using the 
$\beta$--profile (as in the previous case) to simulate the SZ effect. Whereas 
the density profile (Eq. \ref{eq:fit}) is used to simulate the potential well 
of the moving lens effect. We note
that the results of the N-body simulations of
Navarrro, Frenk \& White (1996) are consistent with the assumption of an 
intra--cluster isothermal gas in hydrostatic equilibrium with a NFW halo.
\section{Results of the data analysis}
We analyse the simulated maps of secondary fluctuations due to the moving lens 
effect, for the three cosmological models 
described in Sect. 3, and we quantify their 
contributions. We also make attempts at detecting and extracting the
secondary fluctuations from the entire signal (primary CMB, SZ kinetic effect 
and moving lenses).
\subsection{Statistical analysis}
\begin{figure}
\epsfxsize=\columnwidth
\hbox{\epsffile{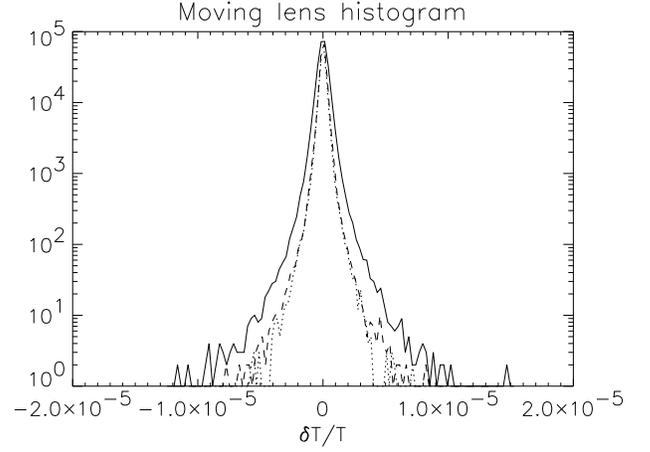}}
\caption{{\small\it Histograms showing the distributions of the secondary 
fluctuations in the simulated maps. The solid, dashed and dotted lines are for
respectively the standard, Open and Lambda CDM model.}}
\label{fig:comphist}
\end{figure}

We show the histogram of the secondary fluctuations for the moving lens effect 
(randomly generated) in the three cosmogonies (Fig. \ref{fig:comphist}). In all cases,
the amplitude of anisotropies ranges roughly between $\delta T/T\simeq -1.5
\,10^{-5}$ and $\delta T/T\simeq 1.5\,10^{-5}$. The 
{\it rms} value of the anisotropies varies a little with the cosmological
model $(\frac{\delta T}{T})_{rms}^{CDM}\simeq5.2\,10^{-7}$,
$(\frac{\delta T}{T})_{rms}^{\Lambda CDM}\simeq3.4\,10^{-7}$ and finally
$(\frac{\delta T}{T})_{rms}^{OCDM}\simeq3.5\,10^{-7}$. Our results are in 
general agreement with those of Tuluie, Laguna \& Anninos (1996). In all the cosmological
models, the {\it rms} value of the anisotropies is
about a factor 10 lower than the {\it rms} amplitude of the fluctuations due
to the SZ kinetic effect associated with the same structures, 
which is about $5.\,10^{-6}$; and is about 30 times lower than the 
$\left(\frac{\delta T}{T}\right)_{rms}$
of the primary fluctuations in a standard CDM model.
 The distribution of the
temperature fluctuations induced by moving lenses exhibits a highly non 
Gaussian signature (Fig. \ref{fig:comphist}). The fourth moment of the 
distribution, called the kurtosis, measures the peakedness or flatness of the
distribution relative to the normal one. We find that the
kurtosis for the standard CDM, OCDM and $\Lambda$CDM models are positive and
respectively equal to about 51, 97 and 41. The distributions are thus peaked
(leptokurtics).
\par
\begin{figure}
\epsfxsize=\columnwidth
\hbox{\epsffile{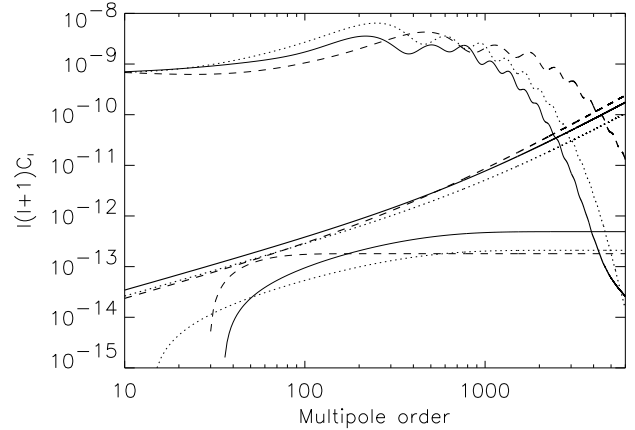}}
\caption{{\small\it Power spectra of the primary fluctuations 
obtained using the CMBFAST code compared to the fitted power 
spectra of the secondary fluctuations due to the Sunyaev--Zel'dovich
kinetic effect (thick lines) and to the moving lens effect (thin lines). The 
power spectra for the 
standard CDM model (solid line), open CDM model (dashed line)
and lambda CDM model (dotted line) are shown.}}
\label{fig:pspec}
\end{figure}

In the context of our statistical analysis of the secondary anisotropies, we
also compute the fitted angular power spectra (Fig. \ref{fig:pspec}) of the three 
main sources of 
anisotropies: primary CMB fluctuations (in the standard CDM model) and both the
predicted power spectra of the fluctuations due to the 
moving lenses (thin lines) and the SZ kinetic effect (thick 
lines). In figure \ref{fig:pspec}, the solid lines are for the standard CDM 
model, dashed and dotted lines are respectively for the open and non zero 
cosmological constant models. 
We fit the power spectra of the secondary anisotropies due to moving lenses
with the general expression:
\begin{equation}
l(l+1)C_l=a_{ls}-b_{ls}\exp(-c_{ls}l),
\end{equation}
in which the fitting parameters for every cosmological model are given in 
table \ref{tab:lspar}.
\begin{table}
\caption{{\small\it Fitting parameters for the power spectrum of the 
fluctuations induced by moving lenses as a function of the cosmological model.
}}
\begin{center}
\begin{tabular}{|c|c|c|c|}
\hline
	& $a_{ls}$ & $b_{ls}$ & $c_{ls}$ \\
\hline
SCDM 	& $4.9\,10^{-13}$ & $5.5\,10^{-13}$ & $3.3\,10^{-3}$ \\
\hline
OCDM	& $1.8\,10^{-13}$ & $7.6\,10^{-13}$ & $4.9\,10^{-2}$ \\
\hline
$\Lambda$CDM & $2.1\,10^{-13}$ & $2.2\,10^{-13}$ & $3.4\,10^{-3}$ \\
\hline
\end{tabular}
\end{center}
\label{tab:lspar}
\end{table}
The SZ kinetic anisotropies are fitted with the following expression:
\begin{equation}
l(l+1)C_l=a_{SZ}\,l+b_{SZ}\,l^2,
\end{equation}
with the fitting parameters for the cosmological models gathered in table.
\ref{tab:szpar}.
\par
\begin{table}
\caption{{\small\it Fitting parameters for the power spectrum of the 
fluctuations induced by the Sunyaev--Zel'dovich kinetic effect as a function of
the cosmological model.}}
\begin{center}
\begin{tabular}{|c|c|c|}
\hline
	& $a_{SZ}$ & $b_{SZ}$ \\
\hline
SCDM 	& $3.4\,10^{-15}$ & $4.3\,10^{-18}$ \\
\hline
OCDM	& $2.3\,10^{-15}$ & $6.3\,10^{-18}$ \\
\hline

$\Lambda$CDM & $2.6\,10^{-15}$ & $2.5\,10^{-18}$ \\
\hline
\end{tabular}
\end{center}
\label{tab:szpar}
\end{table}
\par\medskip
The power spectra of the SZ kinetic effect exhibit the characteristic $l^2$
dependence on small angular scales for the point--like source dominated 
signal. All the power spectra have rather similar amplitudes, at large scales, 
in particular up
to $l\simeq 200$ where we notice an excess of power at small angular scales
in the OCDM model. This is because low $\Omega_0$ models produce higher counts
than $\Omega_0=1$ models (Barbosa et al. 1996).
\par
The moving lens power spectra, for both CDM and $\Lambda$CDM 
models, exhibit a plateau at $l>500$ with a
decrease at larger angular scales. For the OCDM model, the dependence is
roughly constant at all scales. We also note that the highest and lowest power
are obtained, at small angular scales, for respectively the standard CDM and
OCDM models. At large scales, the opposite is true. \\
In order to interpret
this behaviour, we
distinguish between what we refer to as the resolved and unresolved structures.
The spatial extent of the resolved structures is much greater than 
the pixel size (or analogously the beam size). Whereas,  the unresolved objects have
extents close to, or smaller than, the pixel size. At the pixel size an unresolved
structure generates a SZ kinetic anisotropy which is averaged to a non-zero 
value. Whereas the dipolar anisotropy induced by the moving lens effect is
averaged to zero (except what remains from the side effects). A pixel 
size anisotropy thus does not contribute to the signal in the moving lens 
effect;
while it contributes with its $\delta T/T$ amplitude in the SZ kinetic effect.
As a result, the
distribution of the moving lens anisotropies does not reflect the whole
population of objects, but only the distribution of the resolved ones. 
In the OCDM model the structures are more numerous
and form earlier than in a standard CDM model. Consequently, the distribution of 
unresolved objects in OCDM thus shows a large excess compared with the standard
CDM and there are less resolved structures in the OCDM model than in
the CDM. The excess of power in the the moving lens fluctuations spectrum 
(Fig. \ref{fig:pspec}, solid line)
reflects the dependence of the size distribution upon the cosmological model.
\\
At a given large scale and for the SZ kinetic effect, there is more power on 
large scales in a standard CDM model compared with the OCDM. This is because
the contribution to the 
power comes from low redshift resolved structures, which are less numerous in an OCDM
model. Consequently, in the case of the fluctuations induced by the
moving lens effect at large scale, the power in the OCDM model is greater than in the standard
CDM. In addition, 
at a given large scale the power of the moving lens effect accounts for the 
cumulative contribution from the massive objects, with high amplitude, and 
from the less massive ones, with lower amplitudes.
%
%
\par
A comparison between the CMB and the moving lens power spectra obviously shows 
that primary CMB fluctuations dominate at all scales larger than the cut-off
scale, whatever the cosmological model (Fig. \ref{fig:pspec}). 
Furthermore in the OCDM 
and $\Lambda$CDM models the cut-off is shifted towards smaller angular
scales making the CMB the dominant contribution over a larger range of scales.
The most favourable configuration to study and analyse the fluctuations is 
therefore the CDM 
model since it gives the largest cut-off scale compared to the other
cosmological models and since it gives the highest prediction for the power of 
the moving lens effect. 
The level
of spurious additional signal associated with the moving lens effect is 
negligible compared to both the primary and SZ kinetic fluctuations. Below
the scale of the cut-off in the CMB power spectrum, the $l^2$ dependence of the
SZ fluctuations is dominant over the moving lens effect. Moreover,
contrary to the thermal effect, the SZ kinetic, moving lens and primary
fluctuations have black body spectra. This makes the spectral confusion between
them a crucial problem. At small angular scales, the SZ kinetic effect 
represents the principal source of confusion.
\par
\begin{figure}
\epsfxsize=\columnwidth
\hbox{\epsffile{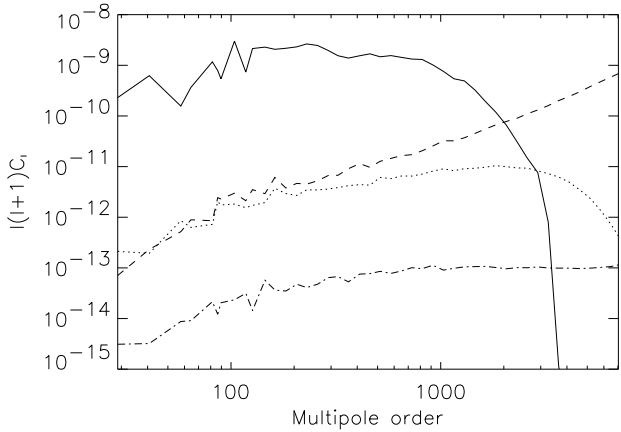}}
\caption{{\small\it Randomly generated power spectra of the primary fluctuations in the standard
CDM model (solid line) compared to the secondary fluctuations due to the 
moving lens effect (dashed-dotted line) and the Sunyaev--Zel'dovich kinetic 
effect with a cut-off at
$10^{13}\,M_{\sun}$ (dashed line) and at $10^{14}\,M_{\sun}$ (dotted line).}}
\label{fig:pspecut}
\end{figure}

Nevertheless, 
the contribution of the SZ kinetic effect is very dependent on the predicted
number of structures that show a gas component. In other words, some objects 
like small groups of galaxies may not have a gas component, and therefore
no SZ thermal or kinetic anisotropy is generated, but they still exhibit the
anisotropy associated with their motion across the sky. 
We attempt to study
a rather wide range of models. We therefore use two prescriptions to 
discriminate 
between ``gaseous'' objects and ``non gaseous'' ones. These prescriptions 
correspond to arbitrary limits on the masses of the structures.
Namely: in the first model, we assume that all the dark matter
halos with masses greater than $10^{13}\,M_{\sun}$ have a gas fraction of 20\%
and exhibit SZ thermal and kinetic anisotropies; while in the second model, it 
is only the structures with masses
$M\geq10^{14}\,M_{\sun}$ which produce SZ anisotropies.
We ran the simulations
with both assumptions in the standard CDM model and computed the corresponding 
power spectra (Fig. \ref{fig:pspecut}). The power spectrum associated with the 
SZ kinetic effect shows, as expected, that the cut-off in masses induces a
decrease in the power of the SZ kinetic effect on all scales, and in
particular on very small scales with a cut-off at $l\simeq4000$. The power 
spectrum of the SZ kinetic anisotropies can be fitted with the following 
expression:
\begin{equation}
l(l+1)C_l=-3.3\,10^{-13}+1.6\,10^{-14}l\exp(6.2\,10^{-4}l).
\end{equation}
Despite this cut-off in mass and the decrease in power, the SZ kinetic
effect remains much larger than the moving lens effect. Therefore at small 
angular scales,
the SZ kinetic point like sources are still the major source of confusion. In 
order to get rid of this pollution in an effective way,
one would need a very sharp but unrealistic cut-off in mass.
\par

\subsection{Detection and extraction}
\begin{figure}
\epsfxsize=\columnwidth
\hbox{\epsffile{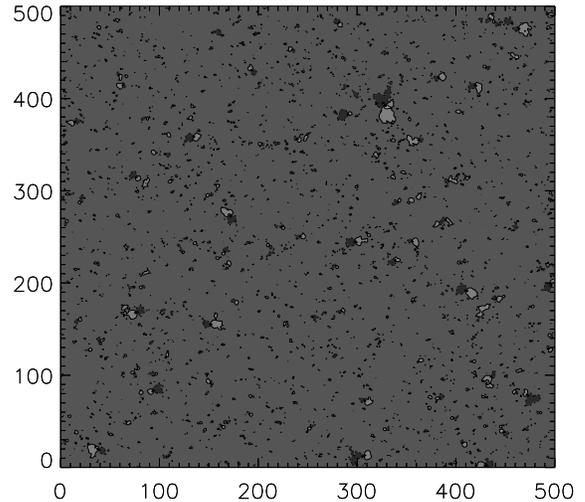}}
\caption{{\small\it Plot showing the contours superimposed over 
a simulated map of the 
fluctuations induced by moving lenses in the standard CDM model 
(pixel size=
1.5'). The contour levels and grey scales shown in the plot are: $1.5\,10^{-5}, \pm1.\,10^{-5},
\pm5.\,10^{-6}, \pm1.\,10^{-6}$}}
\label{fig:contoscdm}
\end{figure}
We analyse the simulated maps in order to estimate the amplitudes of the
anisotropies associated with each individual moving structure. In such an 
analysis both
primary CMB and SZ kinetic fluctuations represent spurious signals with
regards to the moving lens. Figure \ref{fig:pspec} shows that these signals 
contribute at different scales and at different levels. The primary CMB 
contribution vanishes on scales lower than the cut-off whereas the SZ kinetic
contribution shows up at all scales and its power increases as $l^2$ on small
scales. This indicates clearly that the most important problem with the 
analysis
of the maps (extraction and detection of the moving lens anisotropy) is the 
confusion due to the point--like
sources. This problem is made worse by spectral confusion. A compromise must be found between investigating scales 
smaller than the CMB cut-off, which 
maximises the pollution due to SZ kinetic effect,
and exploring larger scales where the SZ contribution is low
(but still 10 times larger than the moving lenses). The main problem here is that
on these scales the primary 
fluctuations are 100 times larger than the moving lenses which makes their 
detection hopeless. 
\par
Nevertheless, the signal has two characteristics that make the attempts at
detection worthy at small scales. The first 
advantage is that the 
anisotropy induced by a moving lens exhibits a particular spatial signature which is seen as the 
dipole--like patterns shown in figure \ref{fig:contoscdm}. The second, 
and main advantage is that we know 
the position of the center of the 
structures  thanks to the SZ thermal effect.
\\
In fact, the objects giving rise to a dipole--like anisotropy are either
small groups or clusters of galaxies with hot ionised gas which 
also exhibit SZ thermal distortions.
The latter, characterised by the so-called Comptonisation 
parameter $y$, have a very specific spectral signature. It is therefore rather
easy to determine the position of the center of a structure assuming that
it corresponds to the maximum value of the $y$ parameter. In the context of
the Planck multi--wavelength experiment for CMB observations, it was shown 
(Aghanim et al. 1997) that the location of massive clusters will be well
known because of the presence of the SZ thermal effect.\\
We based our detection strategy for the moving lens effect on these
two properties (spatial signature and known location). We also assumed that the
SZ thermal effect was perfectly separated from the other contributions thanks 
to the spectral signature. The problem is therefore eased since it lies in the 
separation of moving lens, SZ kinetic and primary CMB anisotropies {\it at
known positions}. Nevertheless the clusters and their gravitational potential 
wells are likely
to be non-spherical, making the separation difficult. In the following, 
we will show that even in the simple spherical model we adopt 
the separation remains very difficult because of the spectral confusion of the
moving lens, SZ kinetic and primary CMB fluctuations. Separation is even more 
difficult because of the numerous point-like SZ kinetic sources corresponding 
to weak clusters and small groups of galaxies for which we do not observe the 
SZ
thermal effect.\\
\subsubsection{Method}
In order to clean the maps from the noise (SZ kinetic and CMB fluctuations), 
we filter them using a wavelet transform. Wavelet transforms have received 
significant attention recently due to their suitability for a number of 
important signal and image processing tasks. The principle behind the wavelet 
transform, as described by Grossmann \& Morlet (1984), Daubechies (1988) and 
Mallat (1989) is to hierarchically decompose an input image into a series of 
successively lower resolution reference images and associated detail 
images. At each level, the reference image and detail image contain the 
information needed to reconstruct the reference image at the next higher
resolution level. So, what makes the wavelet transform interesting in image
processing is that, unlike Fourier transform, wavelets are quite localised in
space. Simultaneously, like the Fourier transform, wavelets are also quite 
localised in frequency, or more precisely, on characteristic scales. Therefore,
the
multi-scale approach provides an elegant and powerful framework for our image
analysis because the features of interest in an image (dipole pattern) are
generally present at different characteristic scales. Furthermore, the wavelet 
transform performs contemporaneously a hierarchical analysis in both the space
and frequency domains.
\par 
The maps are decomposed in terms of a wavelet basis that
has the best impulse response and lowest shift variance 
among a set of
wavelets that were tested for image compression (Villasenor et al. 1995).
These two characteristics are important if we want to identify the locations
and the amplitudes of the moving lenses. Since the moving lenses induce very
small scale anisotropies compared to the CMB, we filter the largest 
scales in order to separate these two contributions. We note that this also
allows us to separate the contributions due to the large scale SZ kinetic 
sources.
In the following we describe our analysis method, first applied to an 
unrealistic study case and then to a realistic case.
\begin{figure*}
\hbox{\epsffile{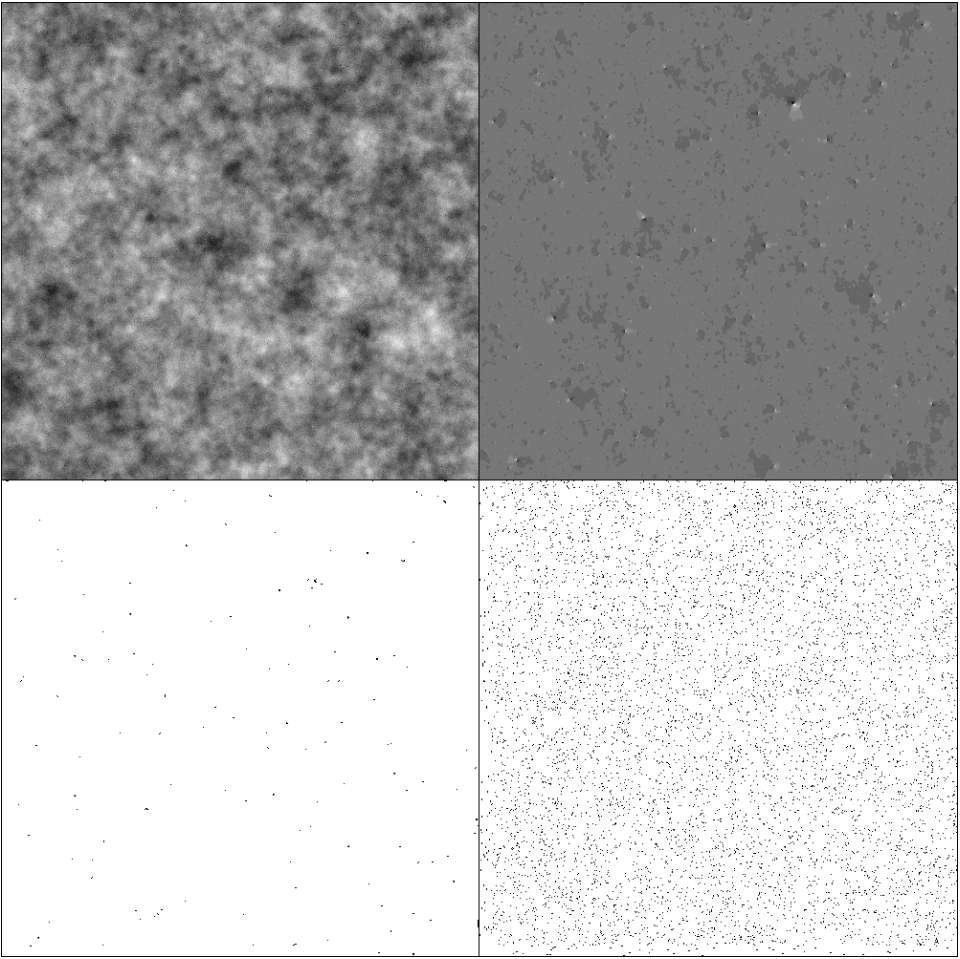}}
\caption{{\small\it The two upper panels show simulated maps: on the left,
total map of the fluctuations (CMB+SZ+lens) linear scale between $1.4\,10^{-4}$
(white) and $-1.3\,10^{-4}$ (black). On the right is shown the map of the 
moving
lenses fluctuations (linear scale between $1.5\,10^{-5}$ (white) and $-1.2\,
10^{-5}$ (black)). The two lower panels are the result of the wavelet 
filtering process. The left panel is for a CMB+lens configuration, in 
which we notice that the secondary anisotropies are rather well extracted.
The right panel represents the case of all contributors and shows that 
the moving lenses are completely dominated by the SZ kinetic noise.}}
\label{fig:impap}
\end{figure*}

\par\noindent
{\bf Study case}
\par
\begin{figure}
\epsfxsize=\columnwidth
\hbox{\epsffile{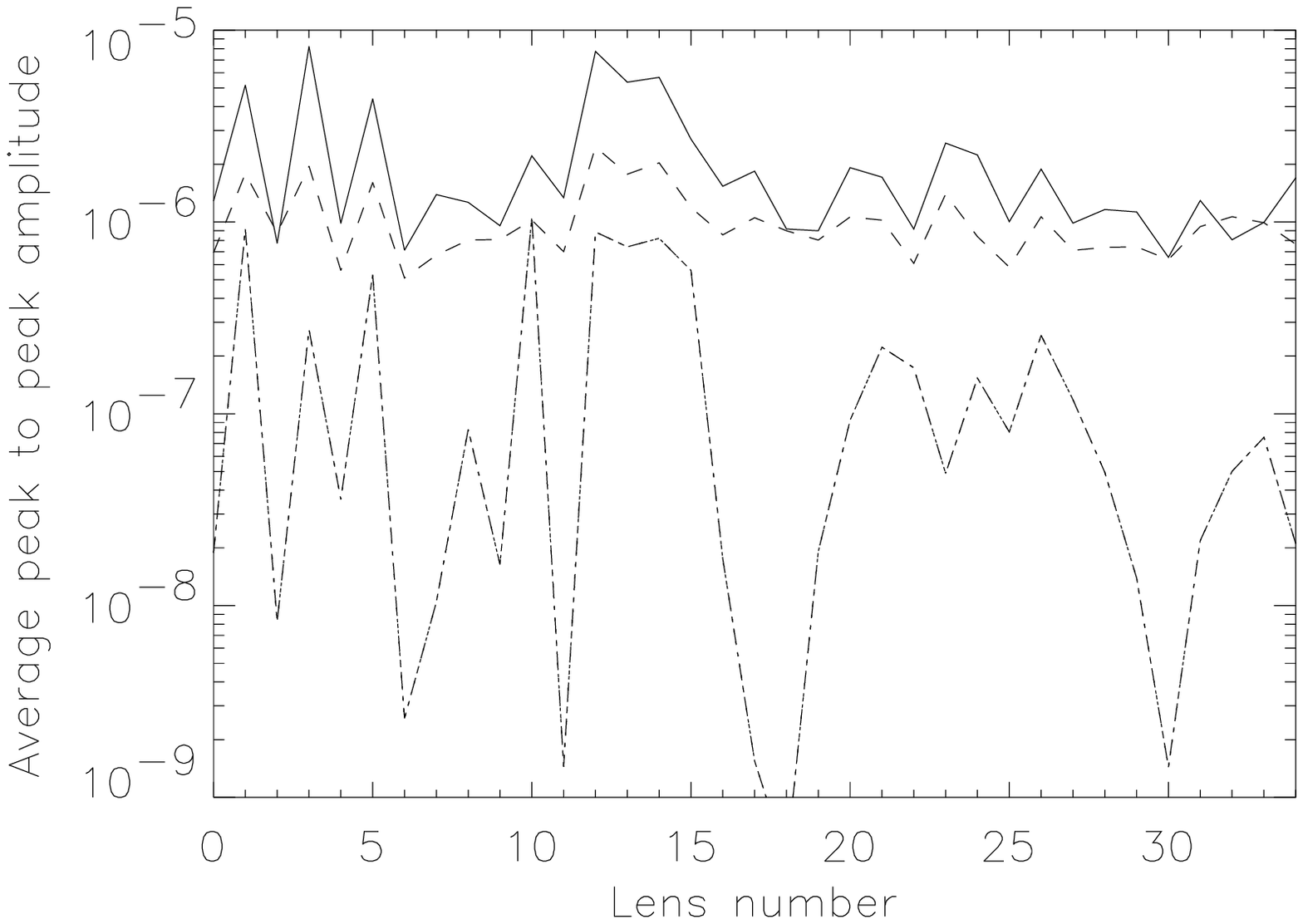}}
\caption{{\small\it  Average peak to peak amplitude of the secondary 
anisotropies due to the moving lenses (cut-off mass $10^{13}\,M_{\sun}$) for
lenses with decreasing $y$ parameter.
Solid line represents the amplitudes in the original simulated lens map.
The dashed line represents the extracted amplitudes 
without sorting the wavelet coefficients and after filtering the 
CBM+lenses contributions (no SZ kinetic). The
dotted-dashed line represents the extracted amplitudes after sorting the 
wavelet coefficients and filtering all contributions (CBM+lenses+SZ 
kinetic).}}
\label{fig:coreltot}
\end{figure}

We filter the large scales of a map of CMB+moving lens fluctuations (no SZ
kinetic contribution) in order to test the robustness and 
efficiency of the wavelet transform filtering. In this case, the noise due to
the CMB is efficiently cleaned. In fact, Figure \ref{fig:impap} lower left 
panel
shows a residual signal (symbolised by the dots) associated with the moving 
lens fluctuations, which are simulated in the upper right panel of the same figure. We
have confirmed that the positions of the residual signal agree with the 
positions
of the input structures. Moreover, we were able to successfully extract the 
secondary fluctuations due to the moving lenses, as well as estimate their 
average peak to peak $\delta T/T$ values. Figure \ref{fig:coreltot} shows the 
average peak to peak amplitudes of the input simulated fluctuations (solid 
line) and the extracted values (dashed line). The main features are 
well-recovered, although the amplitudes suffer from the smoothing of the 
filtering procedure. In this study case, with no SZ kinetic contribution, we
find a correlation coefficient between input and recovered values of about 
0.95.
\par\noindent
{\bf Realistic case}
\par
When this method is applied to filter a map containing all contributions
(CMB + SZ kinetic + moving lenses), we are no longer able to identify or locate 
the moving lens fluctuations, as shown in fig.\ref{fig:impap} lower right 
panel. Here, the CMB which dominates at large scales is cleaned, whereas the
SZ kinetic effect, which is mainly a point-like dominated signal, at least
one order of magnitude larger than the power of the moving lenses, is not 
cleaned and remains in the filtered signal. We have filtered at several 
angular scales 
without any positive result. On large scales the extended dipole patterns are
polluted by the CMB, as mentioned above, and on small scales the SZ kinetic 
fluctuations are of the same scale as the moving lens anisotropies. We also 
tried the convolution of the total map (CMB  + SZ kinetic + moving lenses) with the
dipole pattern function but we were still unable to recover the moving lens
fluctuations. In fact, the combination of two SZ kinetic sources, one coming
forward and the other going backward, mimics a dipole-like pattern. In order
to distinguish between an intrinsic dipole due to a lens and a coincidence, 
one needs to know a priori the direction of the motion which is of course
not possible. During our analysis, we investigated two cases for the cut
off in mass as describe in Sect 4.1. For the simulations with cut-off mass 
$10^{14}\,M_{\sun}$ the resulting background due to
point-like SZ fluctuations is lower than the cut-off at $10^{13}\,M_{\sun}$ case; but
we were still unable to recover the moving lens fluctuations.
\begin{figure}
\epsfxsize=\columnwidth
\hbox{\epsffile{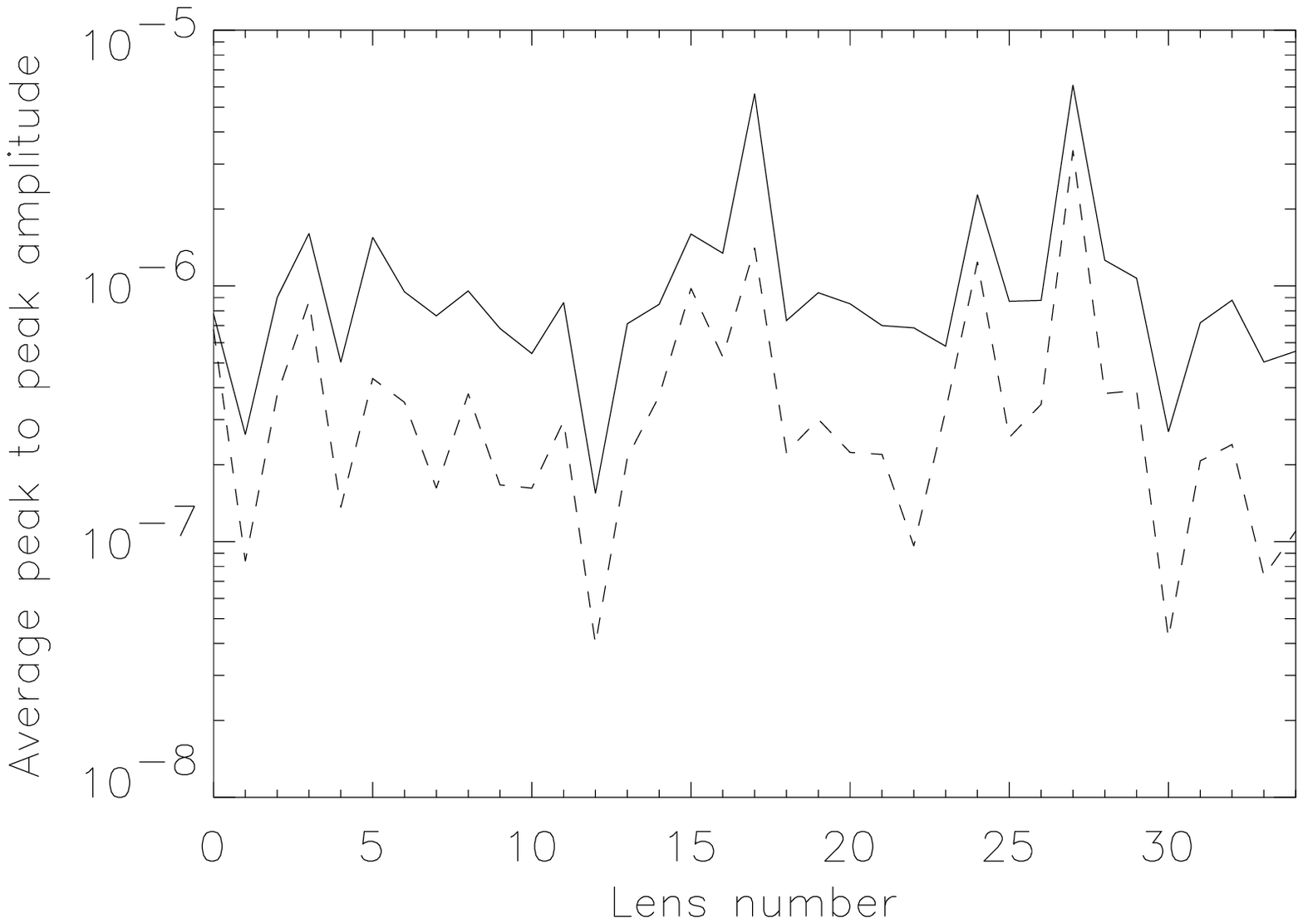}}
\caption{{\small\it Average peak to peak amplitude of the secondary 
anisotropies due to the moving lenses (cut-off mass $10^{14}\,M_{\sun}$) 
for lenses with a decreasing $y$ parameter. The
solid line represents the amplitudes in the original simulated lenses map. The
dashed line represents the extracted amplitudes {\bf after sorting the 
 wavelet coefficients and filtering} all contributions (CMB+SZ+lenses).
 The correlation factor is equal to 0.9.}}
\label{fig:corelcut}
\end{figure}
\par\medskip
In our attempt at taking advantage of the spatial signature of the moving lens 
fluctuations, we have located the coefficients in the wavelet decomposition
that are principally associated with the moving lenses and 
selected them from
all the wavelet coefficients. Our study case procedure is the following. We 
make the wavelet transform for the moving lens fluctuations and, separately, we
also make the transform for the remaining signals (CMB + SZ kinetic).
We locate the wavelet coefficients for
the moving lenses whose absolute values are higher than the absolute values of
CMB+SZ kinetic coefficients. Then,
we select, in the transform of the total fluctuation map (CMB + SZ + lenses), 
the coefficients corresponding to the previously located ones.
Finally we perform the inverse transform on the map (CMB + SZ + lenses)
according to the selected coefficients. When we compare the average peak
to peak amplitudes of the recovered (Fig. \ref{fig:corelcut}: dashed
line and Fig. \ref{fig:coreltot}: dotted-dashed line) and input (Fig. 
\ref{fig:corelcut} and Fig. \ref{fig:coreltot}, solid line) lens fluctuations, 
we find a very good correlation between the amplitudes of the original and the
reconstructed moving lens fluctuations. The correlation factor is of the 
order of 0.7 for CMB + SZ + lenses with a cut-off mass at 
$10^{13}\,M_{\sun}$ and higher
than 0.9 with the cut-off at $10^{14}\,M_{\sun}$. This difference between the
correlation factors is an effect of the cut-off in masses. In fact, for the 
$10^{14}\,M_{\sun}$ cut-off, the filtered maps are cleaner than for the 
$10^{13}\,M_{\sun}$ cut-off. Therefore, in the latter case some of the lenses
have very little or no signature in the wavelet decomposition, hence they are 
not recovered and the correlation factor decreases.
\par\medskip
The results of our study case confirm that the moving lens fluctuations have a 
significant spatial signature in the total signal although their amplitudes are
very low compared with the CMB and SZ fluctuations. However, it is worth 
noting that such a ``good'' result is obtained only because we use sorted 
coefficients from two separated maps, one containing the lens signal and the
other
containing the polluting signals. In a real case, there is no way to separate
the contributions because of the spectral confusion and therefore 
there is no a priori knowledge of the ``right'' coefficients
in the wavelet decomposition. In our
analysis, we tried several sorting criteria for the coefficients but we
could not find a robust and trustworthy criterion to reproducibly 
discriminate between the wavelet coefficients 
belonging to the moving lens fluctuations and the coefficients belonging to
the noise (SZ kinetic and CMB fluctuations). During the analysis, we could not 
overcome the physical limitation corresponding to the presence of sources of 
SZ kinetic anisotropies at the same scale and with amplitudes at least 10 times
higher than the signal (moving lens fluctuations).

\section{Conclusions}
In our work, we investigate the secondary fluctuations induced by moving lenses
with masses ranging from those of groups of galaxies to those of clusters of 
galaxies in a simple way, based on predicted structure counts and simulated 
maps. This method allows us to explore a rather wide range of scales ($>10$
arcseconds) in various cosmological models. The analysis, in terms of angular
power spectra, show the scales for which the primary
fluctuations are dominant (Fig. \ref{fig:pspec}). In the standard and lambda CDM models, the 
primary anisotropies are dominant respectively for scales $l<4000$ and
$l<4500$ whereas in the Open CDM model 
they are dominant for $l<6000$. In practice, it is thus 
impossible to detect the secondary anisotropies due to moving lenses in the
open model. The standard CDM model shows the smallest cut-off scale with an 
intermediate SZ kinetic pollution, compared to the other two models. It is
therefore the ``best case'' framework for making an analysis and 
predicting the detection of fluctuations and the contributions that they
induce. One must keep in mind that the results quoted in this particular case
represent the ``best'' results we get from the analysis.
\par
The results of our analysis are obtained under the assumption of a
universe that never 
re-ionises, which is of course not the case. The re-ionisation, if it is 
homogeneous, is supposed to somewhat
ease the task of extraction of the pattern. In fact, its main effect is to
damp the angular power spectrum of the primary anisotropies on small scales, 
shifting the cut-off towards
larger scales. In this case, the effect of moving lenses dominates over the 
CMB fluctuations, and the SZ kinetic is not as high as it is on very small
scales. However, if the re-ionisation is late and inhomogeneous, it generates 
additional SZ kinetic-type secondary fluctuations (Aghanim et al. 1996) 
without damping the power spectrum by more than a few percent. 
Here, the re-ionisation might worsen the analysis at small scales. In any 
case, there could be some other additional secondary fluctuations principally
due to the Vishniac effect, that arise in a re-ionised universe.
Our work thus gives a ``best case'' configuration of the problem, with all 
other effects tending to worsen the situation.
\par
We found that the secondary fluctuations induced by the moving 
gravitational lenses can be as high as $1.5\,10^{-5}$; 
with {\it rms} contributions of about 5 to $3.\,10^{-7}$ in the three cosmological models. Even if the moving lens fluctuations have a particular
dipolar pattern and even if they are ``perfectly'' located through their SZ
thermal effect, the detection of the moving lens effect and its separation
from the SZ kinetic and primary fluctuations are very difficult because of the 
very high
level of confusion, on the scales of interest, with the point--like SZ kinetic 
anisotropies and because of spectral confusion. 
\par
We nevertheless analysed the simulated maps using an 
adap\-ted wavelet technique in order to extract the moving lens fluctuations.
We conclude that {\it the contribution of 
the secondary anisotropies due to the moving lenses is thus negligible 
whatever the
cosmological model. Therefore it will not affect the future CMB measurements
except as a background contribution. We have highlighted the 
fact that the moving lens fluctuations have a very
significant spatial signature but we did not 
succeed in separating this contribution from the other signals.}

%
%
%

\begin{acknowledgements}
The authors wish to thank J.-L. Puget for many suggestions and fruitful 
discussions. They wish to thank the referee, M. Birkinshaw, for his helpful
comments that much improved the paper.
The authors thank J.F. Navarro, C. Frenk and S.D. White for kindly providing 
us a FORTRAN routine, computing
the concentrations and the critical densities of the dark matter profiles, and
J.R. Bond for providing the CMB map used in the analysis. The power spectra of 
the primary fluctuations were performed using the CMBFAST code (M. Zaldarriaga 
\& U. Seljak). In addition, we
thank F. Bernardeau, F.-X. D\'esert, Y. Mellier and J. Silk for 
helpful discussions and A. Jones for his careful reading of the paper. 
\end{acknowledgements}
%
%
%

%
\end{document}